\documentclass[12pt]{iopart}
\usepackage{iopams}
\usepackage{graphicx}

\begin{document}

\title{Lattice thermal conductivity of graphene with conventionally isotopic defects}

\author{Vadym Adamyan and Vladimir Zavalniuk}
\address{Department of Theoretical Physics, Odessa I I Mechnikov National University,\\2 Dvoryanskaya St., Odessa 65026, Ukraine}
\eads{\mailto{vadamyan@onu.edu.ua}, \mailto{vzavalnyuk@onu.edu.ua}}

\begin{abstract}
The thermal conductivity of doped graphene flake of finite size is investigated with emphasis on the influence of mass of substituting atoms on this property.
It is shown that the graphene doping by small concentrations of relatively heavy atoms results in a disproportionately impressive drop of lattice thermal conductivity.
\end{abstract}

\pacs{63.22.Rc, 65.80.Ck, 66.70.-f,   61.72.J-, 63.20.kp}

\maketitle

\section{Introduction}
This paper was motivated by recent investigations \cite{BalandinExp1,BalandinExp2,BalandinBulk,BalandinReview,Murali,Jauregui,Chen,ChenBalandin,BalandinReview2012}, in which a real possibility of the experimental measurement of the thermal conductivity of graphene was established and its extremely large value was confirmed.

The proposed and performed experiments were based there on the effect of the temperature shift of the Raman G peak position due to the anharmonicity of graphene lattice \cite{Calizo,Calizo2}. Since that shift may be measured with sufficient accuracy, the thermal conductivity of laser heated graphene may be restored using the heat equation with known sources and given geometrical parameters of the graphene flake. Bearing in mind the efficacy of existing experimental methods we consider in this paper the influence of moderate concentrations of different point defects (carbon isotopes, substituting atoms and chemically adsorbed molecules) on the thermal conductivity of graphene.

Note that real and numerical experiments showed that graphene can be doped if required with boron, nitrogen \cite{DefectsNitrogen5p,DefectsNitrogen20p}, aluminium \cite{DefectsAL12p} etc with concentrations reaching as much as several per cent without breaking the graphene planar structure \cite{FreeMonolayers,DefectsAmb}.

The main task of this work is the elucidation of how the thermal conductivity of doped graphene depends on the temperature and mass of defects using as a tool classical methods of the quantum theory of solids. In doing so we restrict ourselves by modeling all the substituting atoms as isotopic defects, assuming that the account of bond weakening may only give an additional reduction of lattice thermal conductivity \cite{Ziman}.

This paper is organized as follows. In Sections 2 we discuss the main mechanisms of the thermal resistance in graphene.
In Sections 3 and 4  we describe briefly the effect of phonon-phonon scattering and edge scattering as the dominant thermal resistance mechanisms in the case of ideal graphene. In Section 5 the explicit expression for the phonon relaxation time determined by phonon-defect scattering is deduced as a function of the quasi-momentum for each phonon branch. In the last section we summarize the obtained results in the form of a fitting formula for the thermal conductivity of a graphene flake as function of temperature, flake size, mass and concentration of dopant.

\section{Heat transport in graphene and mechanisms of thermal resistance}

The heat carriers in graphene are the phonons and conductive electrons. The ideal graphene is a semi-metal with massless charge carriers. As follows, electron and phonon contributions to the heat transport should be in the ratio $\sim c_{ph.}/c_{el.}$ where $c_{ph.}$ and $c_{el.}$ are characteristic velocities of acoustic phonons and electrons (holes) in graphene. Due to the near 50-fold difference in velocities of electrons and phonons, the electron contribution to the ideal graphene total thermal conductivity appears to be less than 1\% \cite{BalandinExp2}. Besides in semi-metals the number of carriers at low and room temperatures is small compared to that of lattice atoms. Since the phonon scattering on free carriers is proportional to their concentration, then the effect of phonon-electron scattering on the lattice heat transport is small comparing to that of phonon-phonon and phonon-defect scattering \cite{Ziman}. Therefore further in this paper we will assume that the electronic subsystem counts very little in the heat transfer as a whole and can be ignored, the more so that the error of present measurements of graphene heat conductivity is $\gtrsim\!10$\%. With this assumption the finite value of the thermal conductivity of real graphene results from phonon scattering on other phonons, flake edges and point defects.

Phonon-phonon scattering is conditioned only by the anharmonic contributions to the potential energy of inter-atomic interaction. The quasi-momentum and energy conservation laws only allow phonon scattering processes in which at least three phonons are involved. As the probability of collisions of four or more phonons is rather small, especially at low temperatures where the equilibrium number of phonons is not great, their contribution to the thermal resistance may be ignored. Remind that multi-phonon scattering processes are divided into normal processes (N-processes) with the strict conservation of momentum and so-called umklapp processes (U-processes), for which the conservation of total quasi-momentum holds up to a non-zero reciprocal lattice vector. The normal processes do not change the total heat flux, thus they drop out from possible mechanisms of thermal resistance, while the U-processes contribute, in an important manner, to the thermal resistance of semiconductors and isolators at high temperatures \cite{Ziman}. On the other hand, with decreasing temperature the contribution of U-processes to the thermal resistance of graphene is suppressed by effects of the defect-induced and edge phonon scattering.

Considering phonon-defect scattering, we restrict ourselves to study only the impact of isotopic defects on the heat transfer in graphene. Note that along with $\,^{13}\textrm{C}$ isotopes (present in natural carbon at a concentration of $\sim 1\%$) the chemically adsorbed atoms and substitutional atoms (like boron, nitrogen,  aluminium etc) can be to some extent treated as isotopic defects. In reality, concentrations of all mentioned defects may range from zero up to several per cent. However, as we show below,  for the low atomic mass of carbon even a concentration of heavy "isotopic" defects $\lesssim 1$\% may double the thermal resistance.

At low temperatures the intensities of all mentioned processes rapidly decrease and the edge scattering is thus rendered the main mechanism limiting the heat transport. Taking account of the very high Debye temperature of graphene and exponential drop in the probability of U-processes there is a temperature range within which only edge and defect scattering should be considered.

Taking into account all the above mentioned contributions, the total relaxation time of a certain phonon state is given by
\begin{equation}\label{TauTotal}
\frac{1}{\tau_{\mathbf{k},s}} = \frac{1}{\tau^{U}_{\mathbf{k},s}} + \frac{1}{\tau^{def}_{\mathbf{k},s}} + \frac{1}{\tau^{edge}_{\mathbf{k},s}},
\end{equation}
where $\mathbf{k}$ is the phonon wave vector and $s$ is the oscillation branch; $\tau^{U}_{\mathbf{k},s}$, $\tau^{def}_{\mathbf{k},s}$ and $\tau^{edge}_{\mathbf{k},s}$ are the relaxation times determined by the phonon-phonon, phonon-defect and edge scattering, respectively.

If the temperature of graphen is well below its Debye temperature (the latter is near 2000K), then the phonon-phonon interaction can be considered as a rather small perturbation and thus different phonon states can be considered as independent. Thus the total lattice thermal conductivity $\kappa$ can be represented as a sum over all possible phonon modes  \cite{Ziman}
\begin{equation}\label{KappaSum}
\kappa = \frac{1}{2} \sum\limits_{\mathbf{k},s} c_{\mathbf{k},s} v_{\mathbf{k},s} \Lambda_{\mathbf{k},s}
 = \frac{1}{2} \sum\limits_{\mathbf{k},s} c_{\mathbf{k},s} v^2_{\mathbf{k},s} \tau_{\mathbf{k},s} ,
\end{equation}
where $\mathbf{k}$ runs the first Brillouin zone, $c_{\mathbf{k},s}$ is the specific heat per normal mode,
\[c_{\mathbf{k},s}=k_{B}\left(\frac{\hbar \omega_{s}(\mathbf{k})}{2 k_{B} T}\right)^{2} \frac{1}{\sinh^2\left(\frac{\hbar \omega_{s}(\mathbf{k})}{2 k_{B} T}\right)},\]
where $s$ numbers the phonon branches, $\omega_{\mathbf{k},s}$ is the frequency of the $|\mathbf{k},s\rangle$-phonon,
$\mathbf{v}_{\mathbf{k},s}=\vec{\nabla}_{\mathbf{k}} \omega_{\mathbf{k},s}$ is the phonon group velocity and $\tau_{\mathbf{k},s}$ is defined by (\ref{TauTotal}).

Since our main task is to study the impact of defects on the thermal conductivity of ideal graphene, we will not scrutinize the contribution of the phonon-phonon U-processes using the same general expression as in \cite{BalandinExp2, BalandinReview, Klemens, BalandinPhonons}; however, with $\omega_{s}(\mathbf{k})$ defined as described below (\ref{eigen}):
\begin{equation}\label{TauU}
\tau^U_{\mathbf{k},s}(T) =\frac{m \overline{v_s}^2}{\gamma_{s}^2} \frac{1}{k_{B} T} \frac{\omega_{D,s}}{\omega_{s}^2(\mathbf{k})}, \quad s=1,\,2,\,3 \,\, ,
\end{equation}
where $m$ is the atomic mass of $\,^{12}\textrm{C}$ isotope and  $\omega_{D,s}$ are conventional Debye frequencies for three acoustic phonon modes ($\omega_{D,1}=2.66\times10^{14}$, $\omega_{D,2}=2.38\times10^{14}$ and $\omega_{D,3}=1.32\times10^{14}$ rad/s).

In order to find from (\ref{KappaSum}) certain values of $\kappa$  we use the explicit expressions for $\omega_{\mathbf{k},s}$ in graphene obtained in \cite{Adamyan} on the basis of the following computationally simple harmonic nearest neighbor model.

Considering the total potential energy $W$ of ideal graphene as a sum of three components depending on different types of interaction between the nearest neighbors:

 - in-plane central forces with a force constant $J_{1}$;

 - in-plane non-central forces with a force constant $J_{2}$;

\begin{figure}[!hbp]
\includegraphics[scale=1]{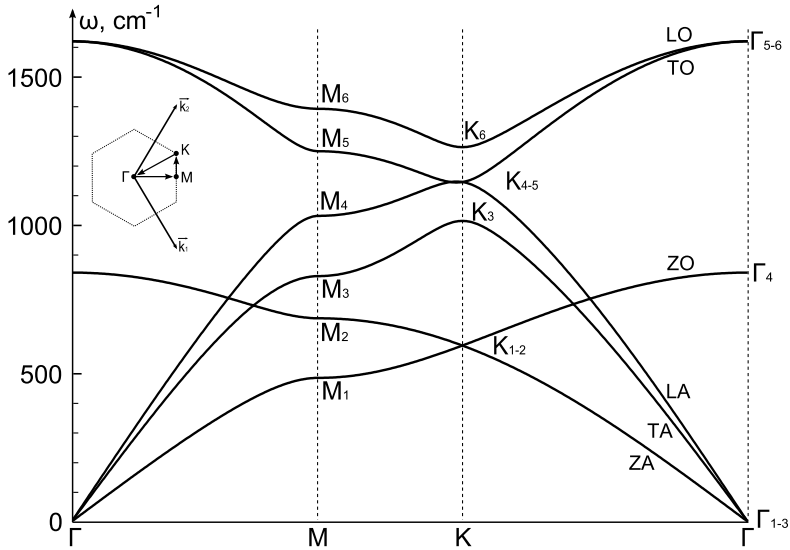}
\caption{Phonon frequency dispersion curves of infinite ideal graphene along $\Gamma-\mathrm{M}-\mathrm{K}-\Gamma$ directions. LA (LO), TA (TO) and ZA (ZO) are longitudinal, transversal and out-of-plane acoustical (optical) branches respectively.}\label{PhononDisp}
\end{figure}

 - out-of-plane non-central forces conditioned by $\pi$-electrons with a force constant $J_{3},$
\\
and applying Bloch's theorem one can reduce within the framework of this model the problem of finding the sought frequencies $\omega_{s}(\mathbf{k})$ to evaluation of eigenvalues of the ideal graphene dynamical matrix
\begin{equation}\label{DynamicalMatrix}
 \mathbf{D}(\mathbf{k})=\frac{1}{m_{0}}\left(\begin{array}{cc}
                     \mathbf{D}^{d}   & \mathbf{D}^{a}  \\
                     \overline{\mathbf{D}^{a}}   &  \mathbf{D}^{d}
                   \end{array}
             \right)
\end{equation}
where
\[\mathbf{D}^{d} =\frac{3}{2}
                             \left(\begin{array}{ccc}
                              J_{1}\!+\!2 J_{2}   & 0 & 0 \\
                              0 & J_{1}\!+\!2 J_{2}  & 0 \\
                              0   &  0  & 2 J_{3}
                             \end{array}\right),
\]
\[\mathbf{D}^{a} =-\frac{J_{1}}{2}
                             \left(\begin{array}{ccc}
                              1\!+\!2 \rme^{\rmi k_1 a}  & 1\!-\!\rme^{\rmi k_2 a} & 0 \\
                              1\!-\!\rme^{\rmi k_1 a}  &  1\!+\!2 \rme^{\rmi k_2 a} & 0 \\
                              0   &  0  & 0
                             \end{array}\right)
 - (1+\rme^{\rmi k_1 a}+\rme^{\rmi k_2 a})
                             \left(\begin{array}{ccc}
                              J_{2} & 0 & 0 \\
                              0 &  J_{2} & 0 \\
                              0   &  0  & J_{3}
                             \end{array}\right).
\]
In this way we find three acoustic and three optical branches $\omega_{s}(\mathbf{k})$ of phonon spectra  (figure \ref{PhononDisp}).
With notation
\begin{equation}\label{auxil}
\begin{array}{l}
X_0=J_1 J_2 + J_2^2,  \qquad\quad  X_1 = J_1^2 + 16 X_0,  \qquad\quad  X_2 = J_1^2 - 8 X_0, \\
F_0(\mathbf{k}) = 2 \left[\cos{(k_1 a \! - \! k_2 a)}+\cos{k_1 a}+\cos{k_2 a}\right], \\
F_1(\mathbf{k}) =  12 (J_1^2 + 2 X_0) +  F_0(\mathbf{k}) (J_1^2 + 8 X_0), \\
F_2(\mathbf{k}) =\left\{9 J_1^2 + X_1 + 2 X_2 \cos{(k_1 a\! - \!k_2 a )}\left(\cos{k_1 a} + \cos{k_2 a}\right) \right. + \\
  \qquad\qquad \left. +  2 \cos{k_1 a} \cos{k_2 a} \left[X_2\! +\! X_1 \cos{(k_1 a\! - \! k_2 a)}\right] - 3 J_1^2 F_0(\mathbf{k}) \right\}^\frac{1}{2},
\end{array}
\end{equation}
the expressions for the eigenfrequencies $\omega_{s}(\mathbf{k})$ take form:
\begin{equation}\label{eigen}
\begin{array}{l}
\omega_{ZA,ZO}(\mathbf{k}) = \left[ \frac{J_3}{m} \left(3 \pm \sqrt{3 + F_0(\mathbf{k})}\right) \right]^{1/2},\\
\omega_{LA,TA,LO,TO}(\mathbf{k}) = \left[ \frac{3 (J_1 + 2 J_2)}{2 m} \pm \frac{\sqrt{2}}{4 m} \sqrt{F_1(\mathbf{k}) \pm 2 J_1 F_2(\mathbf{k})} \right]^{1/2}.
\end{array}
\end{equation}

The values of the force constants
\begin{equation}\label{const}
\begin{array}{l}
J_{1}=135 \rm{\;J\,m^{-2},} \\ J_{2}=245 \rm{\;J\,m^{-2},} \\ J_{3}=83.9 \rm{\;J\,m^{-2}}.
\end{array}
\end{equation}
were chosen in \cite{Adamyan} to fit in the experimental data on inelastic X-ray scattering in graphene \cite{Maultzsch}. It is noteworthy that the model with these $J_{1},\, J_{2},\, J_{3}$ reproduces Raman experiment results and also gives appropriate values for sound velocities
\begin{equation}\label{sound}
\begin{array}{ccc}
v_{LA}=18.4 \rm{\;km\;s^{-1},} & v_{TA}=16.5 \rm{\,km\;s^{-1},} & v_{ZA}=\;\; 9.2 \rm{\,km\;s^{-1}}
\end{array}
\end{equation}\label{const}
in graphene (see \cite{Maultzsch,TA1,TA2}).

We remark that the given expression (\ref{eigen}) for the out-of-plane (bending) mode $\omega_{ZA}(\mathbf{k})$ is inconsistent in the limit $|\mathbf{k}| \rightarrow 0$ with known dependence $\omega_{ZA}(\mathbf{k})\sim k^{2}$. However the quasi-linear dependence of $\omega_{ZA}(\mathbf{k})\simeq v_{ZA} |\mathbf{k}|$ which follows from (\ref{eigen}) occurs in graphene $\omega_{ZA}(\mathbf{k})$-dispersion curves obtained in different ways already for $|\mathbf{k}|>\frac{1}{20}\frac{\pi}{a}\div\frac{1}{10}\frac{\pi}{a}$ or for $\omega_{ZA}\gtrsim2\div4$ meV \cite{BalandinReview,Maultzsch,Disp1,Disp3}. This gives grounds for using the formula (\ref{eigen}) for $T\gtrsim20-30$ K. It is worth recalling that the out-of-plane phonon mode add little to the thermal transport in pure graphene because of its low group velocity and large Gruneisen parameter \cite{BalandinBulk,BalandinReview,BalandinReview2012,Mounet}. By our calculations, the contribution of this mode to the total graphene thermal conductivity at room temperature does not exceed $\approx 15-20\%$.

\begin{figure}[!hbp]
\includegraphics[scale=1]{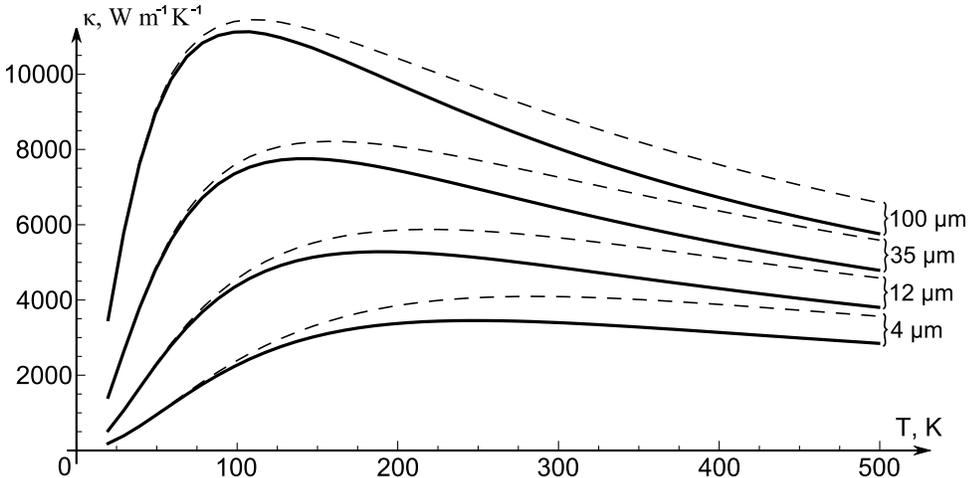}
\caption{Thermal conductivities of ideal graphene flakes of different sizes calculated using the exact expressions (\ref{eigen}) (solid lines) and obtained within the Debye approximation (dashed lines).}\label{TermCondDebye}
\end{figure}

In all our calculations we used the exact expressions (\ref{eigen}) without any further simplification. At low temperatures ($<100$K) this does not improve visibly the Debye approximation. However, the discrepancy between results of the Debye approximation and those obtained using (\ref{eigen}) are quite visible even at room temperature, especially for small graphene flakes (figure \ref{TermCondDebye}).

\section{Effect of the edges}

In the low temperature limit where the U-scattering processes are not active and defect scattering is relatively weak the phonon mean free path $\Lambda$ grows rapidly with $T^{-1}$. On the other hand $\Lambda$ cannot exceed the lattice linear size.  Thus the thermal conductivity in this region is limited only by the edge scattering and should depend linearly on the scaling factor for similar graphene flakes.

For a bounded convex graphene flake with diameter much greater than the lattice constant, the distribution of eigenfrequencies is just the same as that in the case of infinite ideal graphene. The phonon eigenfrequencies of the flake numbered in increasing order are well within the dispersion curves of ideal graphene. Though the displacement eigenvector $|\mathbf{k},s\rangle$ for the infinite ideal graphene lattice is not an eigenvector for a bounded graphene flake, it is, however, the superposition of the flake's displacement eigenvector corresponding to frequencies close to $\omega_{\mathbf{k},s}$.  Therefore the uncertainty  of the exact eigenfrequencies can be reduced to the uncertainty of quasi-momentum $\mathbf{k}$: $\Delta \omega = \frac{\partial \omega}{\partial k}\Delta k$, where $\Delta k  \sim \frac{2\pi}{\widetilde{L}}=\frac{2\pi}{\sqrt{S}}$, $S$ is the flake's area and $\widetilde{L}=\sqrt{S}$ is its characteristic size.

For the phonon wave associated with $|\mathbf{k},s\rangle$ the energy-time uncertainty principle $\tau^{edge}_{\mathbf{k},s} \Delta E \gtrsim \frac{\hbar}{2}$ gives
\[\frac{\hbar}{2\tau^{edge}_{\mathbf{k},s}} \lesssim \Delta E = \hbar\frac{\partial \omega}{\partial k}\Delta k \sim \hbar\frac{\partial \omega}{\partial k} \frac{2\pi}{\widetilde{L}} = 2\pi\hbar \frac{v_{\mathbf{k}}}{\widetilde{L}}.\]

Setting
\begin{equation}\label{FormFactor}
\Lambda^{edge} = \frac{1}{2\pi\,S}\int\limits_{S} dx\,dy \int\limits_{0}^{2\pi} D(x,y,\phi) d\phi = f \sqrt{S},
\end{equation}
where $D(x,y,\phi)$ is the length of the segment inside the flake of ray coming from the point (x,y) and $\phi$-inclined toward the $x$-axis, and $f$ is the dimensionless flake form-factor ($f\!=\!\frac{8}{3\pi^{3/2}}\! \approx\! 0.479$ for the disc and reduces with increasing of the flake diameter) we use further for the $\tau^{edge}_{\mathbf{k}_,s}$ an approximate formula
\begin{equation}\label{TauEdge}
(\tau^{edge}_{\mathbf{k}_,s})^{-1} = \frac{v_{\mathbf{k},s}}{\Lambda^{edge}} = \frac{v_{\mathbf{k},s}}{f \sqrt{S}}.
\end{equation}
in which the ballistic mean free path $\Lambda^{edge}$ absorbs all the peculiarities of the flake geometry.

In the case of graphene with abundant linear structural defects the boundaries of emerging domains became the main source of the edge scattering. If so, to obtain $\widetilde{L}$ one may calculate it as above for a certain domain with some shape and area $S_{D}$ and then average the obtained value over the distribution of the domain shapes and areas and also incorporating in this case the specularity parameters $p$ at the domain boundaries as in \cite{BalandinReview}.

Note that at low temperatures the finite size of a graphene flake is a dominant factor dictating the finiteness of the phonon free path.
Since all the active phonons are acoustic phonons near the $\Gamma$-point where $\omega_s \approx v_s k$, then the thermal conductivity takes the form $\kappa = \frac{1}{2}\Lambda^{edge}\sum\limits_{s} v_{s} C_{s} \sim T^{2}$, where $s$ numbers acoustic phonon branches and $C_s$ are corresponding partial lattice specific heats per unit volume.

\section{Defect scattering}

The mean free path of the $|\mathbf{k},s\rangle$-phonon caused only by its scattering on point defects is defined by the evident expression
\[\Lambda^{def}_{\mathbf{k},s} = \frac{1}{n \, \sigma_{s}(\mathbf{k})},\]
where $n$ is the surface density of point defects and $\sigma_{s}(\mathbf{k})$ is the corresponding phonon total cross-section. Hence
\begin{equation}\label{TauDef}
\frac{1}{\tau^{def}_{\mathbf{k},s}} = n \, \sigma_{s}(\mathbf{k}) \, |\mathbf{v}_{s,\mathbf{k}}|.
\end{equation}

Using the optical theorem and Lippmann-Schwinger formula for calculation of $\sigma_{s}(\mathbf{k})$ in (\ref{TauDef}) yields \cite{Stoneham} (see Chapter 11)
\begin{equation}\label{TauDefIm}
\frac{1}{\tau^{def}_{\mathbf{k},s}} = \frac{\Omega}{\omega_{\mathbf{k},s}} \textrm{Im} \langle \mathbf{k},s|\mathbf{T^{+}}\mathbf{G}\mathbf{T}|\mathbf{k},s\rangle = \frac{\Omega}{\omega_{\mathbf{k},s}} \textrm{Im} \langle \mathbf{k},s|\mathbf{T}|\mathbf{k},s\rangle.
\end{equation}
where $\Omega$ is the area of a two-dimensional graphene lattice,  \[\mathbf{G}=\mathbf{G}(\mathbf{k},\omega^{2}+i\,0)=\left(\mathbf{D}(\mathbf{k})-\mathbf{I}(\omega^2+i\,0)\right)^{-1}\]
is the Green's function of the infinite ideal graphene lattice with the dynamical matrix $\mathbf{D}(\mathbf{k})$ defined above (\ref{DynamicalMatrix}),
and $\mathbf{T}=\mathbf{T}(\omega_{\mathbf{k},s}^{2}+i\,0)$ is the T-matrix,  which connects the Green's function $\widetilde{\mathbf{G}}$ of a defect lattice with that of the ideal lattice: $\widetilde{\mathbf{G}}=\mathbf{G}-\mathbf{G}\mathbf{T}\mathbf{G}$, or $\mathbf{T}=\mathbf{D}(1+\mathbf{G}\mathbf{D})^{-1}$. For an isotopic defect located at the lattice site (0,0) the element of T-matrix with indices $\gamma,j,l'';\gamma',j',l'''$ (where $\gamma$ numbers graphene sublattices, $j$ is associated with spatial components of the atom displacement and $l$ is the lattice vectors) has the form
\begin{equation}\label{TandQMatrix}
\begin{array}{l}
  \mathbf{T}_{\gamma,\gamma'}^{j,j'}(l'',l''') = - \omega^{2}\frac{\Delta m}{m} \delta_{\gamma,0}\:\delta_{l'',0} \:(\mathbf{Q}^{-\!1})^{j,j'}\:\delta_{\gamma',0}\:\delta_{l''',0}, \\
  \\
  (\mathbf{Q}_{\omega^2+i\,0})^{i,j}:=\delta_{i,j} - (\omega^{2}+i\,0)\frac{\Delta m}{m}\sum\limits_{\mathbf{q}} \mathbf{G}^{i,j}_{0,0}(\mathbf{q},\omega^{2}+i\,0) \end{array}
\end{equation}
and
\begin{equation}\label{ImPart}
\begin{array}{c}
  \textrm{Im} \langle\nu, \mathbf{k}|\mathbf{T}(\omega_{\nu}^{2}+i\,0)|\nu, \mathbf{k}\rangle= \\
  \\
  -\frac{1}{N}\frac{\Delta m}{m} \textrm{Im} \left(\omega^{2}_{\nu}+i\,0\right) \sum\limits_{i,j}\left(\mathbf{e}^{i}_{0, \nu}(\mathbf{k})\right)^{\ast} \left(\left[\mathbf{Q}_{\omega^2_{\nu}+i\,0}\right]^{-1}\right)^{i,j}\mathbf{e}^{j}_{0, \nu}(\mathbf{k}).
\end{array}
\end{equation}

The expressions (\ref{TauDefIm}), (\ref{TandQMatrix}), (\ref{ImPart}) were immediately used for the calculation of $(\tau^{def}_{\mathbf{k},s})^{-1}$.

\section{Discussion}

We used directly the relations for the total lattice thermal conductivity (\ref{TauTotal}), (\ref{KappaSum}) together with the explicit expressions discussed above (\ref{TauU}), (\ref{TauEdge}), (\ref{TauDefIm}) and (\ref{eigen}) for $\tau^{U}_{\mathbf{k},s}, \; \tau^{edge}_{\mathbf{k},s}, \; \tau^{def}_{\mathbf{k},s}$ and $\omega_{\mathbf{k},s}$, to calculate the thermal conductivity of graphene flakes in a wide range of linear sizes, temperatures and concentrations of isotopic defects with $\mu=\frac{\delta m}{m}$ varying from $0$ to $2$.

The obtained data on graphene thermal conductivity in the range $T \in (20,500)$K, $L \in (1,100) \;\rm{\mu m}$, $|\mu|<2$ and $n < 0.05$ are reproduced with an accuracy to $\leq1\%$ by the following empirical formula which permits us to obtain the value of the thermal conductivity without re-calculating explicit expressions (\ref{TauU}), (\ref{TauEdge}), (\ref{TauDefIm}) and (\ref{eigen}) (any further improvement of this formula has no sense in view of the approximations laid down into the above theoretical constructions)
\begin{equation}\label{kappaApprox}
\kappa(\Theta,\mathcal{L},\mu,n) = \left(\frac{1}{\kappa_{I}(\Theta,\mathcal{L})} + \frac{\mu^{2}\, n}{\kappa_{D}(\Theta,\mathcal{L},\mu,n)}\right)^{-1},
\end{equation}
where $\kappa_{I}(\Theta,\mathcal{L})$ is the thermal conductivity of an ideal graphene flake, $\Theta=T/\theta$, $\theta=\hbar\,\omega_{D,1} k_{B}^{-1}$ and $\mathcal{L}=\frac{L}{1\,\rm{\mu m}}$ is the characteristic length of the flake in micrometers,
\[\kappa_{I}(\Theta,\mathcal{L}) = \frac{A_1(\mathcal{L})\Theta^{2}}{[\Theta+A_2(\mathcal{L})]^{3}+A_3^2(\mathcal{L})},\]
\[A_1(\mathcal{L})=\frac{1310}{0.573+\mathcal{L}^{-0.45}};  \quad
A_2(\mathcal{L})=\frac{3}{26\,\mathcal{L}^{0.07}} -0.0594;  \quad
A_3(\mathcal{L})=\frac{5}{121\,\mathcal{L}^{0.35}} - 0.005;\]
and
\[\kappa_{D}(\Theta,\mathcal{L},\mu,n)
 = B_{1}(\mathcal{L})\sqrt{\Theta} + B_{2}(\mathcal{L})\mu\,n^{0.53} + B_{3}(\mathcal{L})\]
\[B_1(\mathcal{L}) = -14.9\,\mathcal{L}^{7/13} + \frac{1711}{11-\log\mathcal{L}} - 107,  \quad
B_2(\mathcal{L})=46.5\frac{\mathcal{L}-153}{\mathcal{L}^{0.5}+8}+913,\]
\[B_3(\mathcal{L})=13.6\,\mathcal{L}^{0.42} - 3.7\log\mathcal{L} - 16.8.\]

\begin{figure}[!hbp]
\includegraphics[scale=1.0]{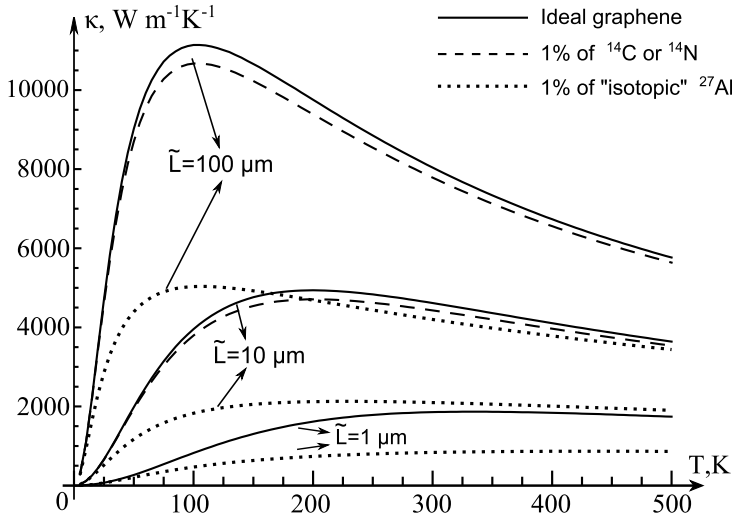}
\caption{
Thermal conductivity for square graphene flakes of different characteristic sizes with aluminium-27 and nitrogen-14 (or carbon-14) isotopic defects.}
\label{TermCond}
\end{figure}

As expected, the results for ideal graphene are close to those of Balandin et al. \cite{BalandinReview}. Notice that the qualitative trend of
the thermal conductivity of doped graphene as a function of temperature, type and concentration of defects emerging from (\ref{kappaApprox}) agrees well with arguments and qualitative conclusions for graphene of the comprehensive review paper \cite{Zhang}. At low temperatures the dominant factor of phonon scattering is the finiteness of a graphene flake that gives $\kappa \sim T^{2} $. For relatively high temperatures the thermal conductivity $\kappa$ varies in inverse proportion to $T$ due to the phonon-phonon scattering. The phonon-phonon scattering is calculated here by summation over all the Brillouin zone in contrast to \cite{BalandinBulk,BalandinReview}, where the Debye approximation was used. The Debye approximation tends to increase the value of thermal conductivity since it gives non-zero sound velocities at the Brillouin zone edges. The difference between these approaches is noticeable for $T\gtrsim100-200$K being as much as 20\% for $T=500$K, but so far remaining within the experimental error.

In order to avoid any additional geometrical and non-geometric parameters, the phonon mean free path associated with the edge scattering is taken simply as (\ref{FormFactor}).

The established decrease of graphene thermal conductivity caused by isotopic defects appears to be approximately linear in concentration $n$ (for small concentrations) and strongly dependent on $\mu$ and temperature. Note that though the cross-section of phonon scattering on defects is by itself temperature independent, it depends substantially on the phonon quasi-momentum. Therefore the relative contributions of phonons from different parts of the first Brillouin zone to the heat transport changes with temperature, resulting in the complex temperature dependence of the contribution of defects to the total heat resistance.

For substitution defects with $|\mu| \leq 1/6$, such as boron, nitrogen and carbon isotopes the relative reduction in thermal conductivity is just proportional to $n \mu^2$ and is well within experimental error. For example, for thermal conductivity of graphene with concentration $n \sim 0.01$ of $\,^{13}$C according to the above results at room temperature the derivative $\frac{\partial\kappa}{\partial n} \sim\!-1 \;\rm{(W m K^{-1})}$ and varies from -1.2 to -0.8 as $L$ increases from 1 to 100 $\mu$m.
For the isotopically modified $\,^{13}$C-graphene with concentration $n\sim0.01$ of $\,^{12}$C this derivative is just the same while the difference in heat conductivities of pure $\,^{12}$C- and $\,^{13}$C-graphene appears to be lesser than that between pure $\,^{12}$C- and pristine graphene. This agrees with experimental results on the thermal conductivity of graphene with concentrations of $\,^{13}$C isotope of 0.011 and 0.992 \cite{Chen}. Furthermore, despite the fact that the model developed above is not valid without further improvement for very high defect concentrations, the ratio of the calculated value of thermal conductivity of pristine graphene to that of 50-50 isotope mixture is the same as in \cite{Chen}. In the meantime we cannot give so far the satisfactory explanation of the drastic drop in the measured value of graphene thermal conductivity for the $\sim\!1\%$ isotope concentrations observed in \cite{Chen}.

For heavier defects (aluminium atoms and different chemically adsorbed molecules) with $\mu \gtrsim 1$ the thermal conductivity drops sufficiently even for $n \sim 1\%$. As an example, 1\% of "isotopic" $\,^{27}$Al ($\mu=1.25$) reduces the thermal conductivity by more than half (figure \ref{TermCond}).
Strictly speaking, the replacement atoms and adsorbed molecules are not isotopic defects. However, the entailed bonds weakening may only enhance the effect of low frequency phonon scattering (see \cite{Ziman}, Chapter VI) and thus lead to a further reduction in the thermal conductivity of graphene.

It should be noted that chemisorbed molecules may desorb in the temperature range 300-600 K decreasing the concentration of defects. So the graphene thermal conductivity under gas-surface equilibrium conditions in corresponding temperature regions may show a sufficient rise with $T$ instead of the "normal" behavior $\kappa \sim T^{-1}$ \footnote{This important comment was advanced by one of our anonymous referees.}.

Thus the insertion of rather low concentrations of rather heavy dopants has a disproportionately great impact on the thermal resistance of graphene. This conclusion is completely in line with results of rather recent molecular dynamics simulations of phonon transport in doped graphene \cite{Zhang, Hao, ZhangH}.

\ack
The authors are grateful to the reviewers for their useful remarks and comments.

This work was supported by the Ukrainian State Fund for Fundamental Researches, grant no. 0112U001739.



\section*{References}
\begin{thebibliography}{35}

\bibitem{BalandinExp1} Balandin A A, Ghosh S, Bao W, Calizo I, Teweldebrhan D, Miao F and Lau C N 2008 Superior thermal conductivity of single-layer graphene {\it Nano Letters} \textbf{8} 902-7.
\bibitem{BalandinExp2} Ghosh S, Calizo I, Teweldebrhan D, Pokatilov E P, Nika D L, Balandin A A, Bao W, Miao F and Lau C N 2008 Extremely high thermal conductivity of graphene: Prospects for thermal management applications in nanoelectronic circuits {\it Appl. Phys. Lett.} \textbf{92} 151911.
\bibitem{BalandinBulk} Nika D L, Ghosh S, Pokatilov E P and Balandin A A 2009 Lattice thermal conductivity of graphene flakes: comparison with bulk graphite {\it Appl. Phys. Lett.} \textbf{94} 203103.
\bibitem{BalandinReview} Balandin A A, Ghosh S, Nika D L and Pokatilov E P 2010 Thermal conduction in suspended graphene layers {\it Fullerenes, Nanotubes, and Carbon Nanostructures} \textbf{18} 474-86.
\bibitem{Murali} Murali R, Yang Y, Brenner K, Beck T and Meindl J D 2009 Breakdown current density of graphene nanoribbons {it Appl. Phys. Lett.} \textbf{94} 243114-1-3.
\bibitem{Jauregui}  Jauregui L A, Yue Y, Sidorov A, Hu J, Yu Q, Lopez J, Jalilian R, Benjaming D K, Delkd D A, Wu W  et al. 2010 Thermal transport in graphene nanostructures: experiments and simulations {\it ECS Transactions} \textbf{28} 73-83.
\bibitem{Chen} Chen S, Moore A L, Cai W, Suk J W, An J, Mishra C at al. 2011 Raman measurements of thermal transport in suspended monolayer graphene of variable sizes in vacuum and gaseous environments {\it ACS Nano} \textbf{5} 321-8.
\bibitem{ChenBalandin} Chen S, Wu Q, Mishra C, Kang J, Zhang H, Cho K, Cai W, Balandin A A and Ruoff R S. 2012 Thermal conductivity of isotopically modified graphene {\it Nature Materials} \textbf{11} 203-7.
\bibitem{BalandinReview2012} Nika D L and  Balandin A A 2012 {\it J.Phys.: Condens. Matter} {\bf 24} 233203.
\bibitem{Calizo} Calizo I, Miao F, Bao W, Lau CN and Balandin A A 2007 Variable temperature Raman microscopy as a nanometrology tool for graphene layers and graphene-based devices {\it Appl. Phys. Lett.} \textbf{91} 071913(1-3).
\bibitem{Calizo2} Calizo I, Balandin A A, Bao W, Miao F and  Lau C N 2007 Temperature Dependence of the Raman Spectra of Graphene and Graphene Multilayers {\it Nano Letters} \textbf{7} 2645-9.
\bibitem{DefectsNitrogen5p} Li X, Wang H, Robinson J T, Sanchez H, Diankov G and Dai H 2009 Simultaneous nitrogen doping and reduction of graphene oxide {\it J. Am. Chem. Soc.} \textbf{131} 15939-44.
\bibitem{DefectsNitrogen20p} dos Santos M C and Alvarez F 1998 Nitrogen substitution of carbon in graphite: structure evolution toward molecular forms \PR B \textbf{58} 13918-24.
\bibitem{DefectsAL12p} Ao Z M, Jiang Q, Zhang R Q, Tan T T and Li S 2009 Al doped graphene: a promising material for hydrogen storage at room temperature \JAP \textbf{105} 074307.
\bibitem{FreeMonolayers} Berciaud S, Ryu S, Brus L E and Heinz T F 2009 Probing the intrinsic properties of exfoliated graphene: Raman spectroscopy of free-standing monolayers {\it Nano Lett.} \textbf{9} 346-52.
\bibitem{DefectsAmb} Casiraghi C, Pisana S, Novoselov K S, Geim A K and Ferrari A C 2007 Raman fingerprint of charged impurities in graphene {\it Appl. Phys. Lett.} \textbf{91} 233108-10.
\bibitem{Ziman}  Ziman J M 1960 {\it Electrons and phonons. The Theory of Transport Phenomena in Solids} (Oxford: Clarendon Press).
\bibitem{Klemens} Klemens P G 2001 Theory of thermal conduction in thin ceramic films {\it International Journal of Thermophysics} \textbf{22} 265-75.
\bibitem{BalandinPhonons} Nika D L, Pokatilov E P, Askerov A S and Balandin A A 2009 Phonon thermal conduction in graphene: role of Umklapp and edge roughness scattering \PR B \textbf{78} 155413.
\bibitem{Adamyan} Adamyan V and Zavalniuk V 2010 Phonons in graphene with point defects. {\it J.Phys.: Condens. Matter} {\bf 23} 015402-1-10.
\bibitem{Maultzsch} Mohr M, Maultzsch J, Dobard\v{z}i\'{c} E, Reich S, Milo\v{s}evi\'{c} I, Damnjanovi\'{c} M, Bosak A, Krisch M and Thomsen C 2007 Phonon dispersion of graphite by inelastic x-ray scattering \PR B \textbf{76} 035439.
\bibitem{TA1} Oshima C, Aizawa T, Souda R, Ishizawa Y and Sumiyoshi Y 1988 Surface phonon dispersion curves of graphite (0 0 0 l) over the entire energy region \SSC \textbf{65} 1601-4.
\bibitem{TA2} Siebentritt S, Pues R, Rieder K-H and Shikin A M 1997 Surface phonon dispersion in graphite and in a lanthanum graphite intercalation compound  \PR B \textbf{55} 7927-34.
\bibitem{Disp1} Wirtz L and Rubio A 2004 The phonon dispersion of graphite revisited  {\it Solid State Communications} \textbf{131} 141-52.
\bibitem{Disp3} Politano A, Marino A R, Campi D, Far\'{\i}as D, Miranda R and Chiarello G 2012  Elastic properties of a macroscopic graphene sample from phonon dispersion measurements {\it Carbon} \textbf{50} 4903-10.
\bibitem{Mounet} Mounet N and Marzari N 2005 First-principles determination of the structural, vibrational and thermodynamic properties of diamond, graphite, and derivatives \PR B \textbf{71} 205214.
\bibitem{Stoneham} Stoneham A M 1975 {\it Theory of defects in solids. Vol. 1} (Oxford: Clarendon Press).
\bibitem{Zhang} Zhang G and Li B 2010 Impacts of doping on thermal and thermoelectric properties of nanomaterials {\it Nanoscale} \textbf{2}, 1058-68.
 \bibitem{Hao} Hao F, Fang D, Xu Z 2011 Mechanical and thermal transport properties of graphene with defects {\it Appl. Phys. Lett.} \textbf{99} 041901.
\bibitem{ZhangH} Zhang H, Lee G, Cho K 2011 Thermal transport in graphene and effects of vacancies \PR B \textbf{84} 115460.

\end{thebibliography}
\end{document}